\documentclass[12pt]{iopart}
\usepackage[latin1]{inputenc}
\usepackage[T1]{fontenc}
\usepackage{ae}
\usepackage[automark]{scrpage2}
\usepackage{array}
\usepackage{amsfonts, amssymb, amsthm, amstext}
\usepackage{graphicx}
\usepackage{hyperref}
\usepackage[usenames, dvipsnames]{color}
\usepackage{subfig}
\usepackage[subfigure]{tocloft}

\newcommand{\prespectrometer}{pre-spec\-tro\-meter}

\begin{document}

\hypersetup{
	pdfsubject = {Modeling of electron emission processes accompanying Radon-$\alpha$-decays within electrostatic spectrometers},
	pdftitle = {Modeling of electron emission processes accompanying Radon-$\alpha$-decays within electrostatic spectrometers},
	pdfauthor = {},
	pdfcreator = {LaTeX, pdfTeX},
	pdfproducer = {KIT},
	pdfkeywords = {KATRIN, background, radon, decay, electron, spectrometer}
}

\title[Electron emission processes accompanying Radon-$\alpha$-decays]{Modeling of electron emission processes accompanying Radon-$\alpha$-decays within electrostatic spectrometers}

\author{N. Wandkowsky$^1$, G. Drexlin$^1$, F.M. Fr{\"a}nkle$^{1,2}$, F. Gl{\"u}ck$^{1,3}$, S. Groh$^1$ and S. Mertens$^{1,4}$}

\address{$^1$ KCETA, Karlsruhe Institute of Technology, 76131 Karlsruhe, Germany}
\address{$^2$ Department of Physics, University of North Carolina, Chapel Hill, NC, USA}
\address{$^3$ Research Institute for Nuclear and Particle Physics, Theory Dep., Budapest, Hungary}
\address{$^4$ Institute for Nuclear \& Particle Astrophysics, Lawrence Berkeley National Laboratory, CA, USA}

\ead{nancy.wandkowsky@kit.edu}


\begin{abstract}
Electrostatic spectrometers utilized in high-resolution $\beta$-spectroscopy studies such as in the Karlsruhe Tritium Neutrino (KATRIN) experiment have to operate with a background level of less than $10^{-2}$~counts per second.
This limit can be exceeded by even a small number of ${}^{219,220}$Rn atoms being emanated into the volume and undergoing $\alpha$-decay there.
In this paper we present a detailed model of the underlying background-generating processes via electron emission by internal conversion, shake-off and relaxation processes in the atomic shells of the ${}^{215,216}$Po daughters.
The model yields electron energy spectra up to 400~keV and electron multiplicities of up to 20 which are compared to experimental data.

\end{abstract}

\maketitle


\section{Introduction}
\label{sec:intro}

The Karlsruhe Tritium Neutrino experiment (KATRIN) is a next generation, large-scale tritium $\beta$-decay experiment designed to determine the effective electron (anti-)neutrino mass $m_{\bar{\nu}_{e}}$ with a sensitivity of $200~\mathrm{meV}$ (90\% C.L.)~\cite{DesignReport}.
It is currently being assembled by an international collaboration at the Karlsruhe Institute of Technology (KIT) in Germany.

KATRIN will investigate the kinematics of tritium $\beta$-decay with unprecedented precision in a narrow region close to the $\beta$-decay endpoint $E_{0}~\approx~18.6$~keV~\cite{Review}. 
It is only in this narrow region with almost vanishing neutrino momenta that one can gain access to $m_{\bar{\nu}_{e}}$. 
An essential pre-requisite to obtain the reference sensitivity of 200~meV is a low background level of $<10^{-2}$~counts per second (cps) in the signal region close to $E_{0}$.

The KATRIN setup is described in detail in~\cite{DesignReport}.
It consists of a windowless gaseous tritium source providing $>10^{11}$ $\beta$-decays per second, a differential and cryogenic pumping section to eliminate the injected tritium molecules from the beam line, as well as an electrostatic spectrometer acting as high-pass energy filter of unprecedented precision, and finally a position sensitive detector to count transmitted electrons.
This work is focused on background processes in the large spectrometer section.

In a previous publication~\cite{Fraenkle} we have reported on measurements with the KATRIN \prespectrometer{} in a test set-up configuration where $\alpha$-decays of ${}^{219,220}$Rn atoms in the volume of an electrostatic spectrometer of the MAC-E filter type\footnote{Magnetic Adiabatic Collimation with Electrostatic filter}~\cite{MACE,MACE1,MACE2} were identified as significant source of background. 
These atoms originate mainly from the non-evaporable getter material which is used as a chemical pump to obtain ultra-high vacuum (UHV) conditions of $p~\leq~10^{-10}$~mbar~\cite{Vacuum}, but also from other auxiliary equipment within the spectrometer vessel and from the stainless steel vessel hull itself. 
In particular, we could demonstrate that a single radon $\alpha$-decay can produce up to several thousands of detector hits in the energy region-of-interest over an extended time period of up to several hours. 
This background originates from the emission of electrons in the energy range from~eV up to several hundreds of~keV, which is caused by a variety of processes related to the emission of the energetic $\alpha$-particle as well as the subsequent reorganization of the atomic shells. 
Almost all of these electrons are trapped in the sensitive volume of the spectrometer due to the known magnetic bottle characteristic of a MAC-E filter~\cite{MagneticMirror,NuclearDecay}. 
Owing to the excellent UHV conditions in this part of the setup, electrons remain trapped over very long time periods, so that they can produce secondary electrons via ionization of residual gas molecules. 
A significant fraction of these secondaries can reach the detector, resulting in a background rate exceeding the KATRIN design limit of $10^{-2}$~cps.

In this paper we describe a detailed model of electron emission processes following $\alpha$-decays of the isotopes ${}^{219,220}$Rn. 
In a separate publication~\cite{RadonValidation} we in detail validate this model experimentally by making use of precise electron trajectory calculations in a MAC-E filter to describe the initial background investigations reported in~\cite{Fraenkle,PhDFraenkle,PhDMertens}, as well as the more in-depth studies performed in~\cite{RadonValidation,PhDNancy}. 
Furthermore, we make use of the model of this work to derive estimates of the background rates and topologies for the final KATRIN set-up in~\cite{NuclearDecay}, while an active background reduction technique concerning trapped electrons is described in~\cite{ECR}.

This paper is organized as follows: Section~\ref{sec:alphadecay} will present in detail the processes related to electron emission during and after ${}^{219,220}$Rn $\alpha$-decay, namely internal conversion, inner shell shake-off, relaxation and atomic shell reorganization. 
The implementation of this model into our simulation software will be discussed in section~\ref{sec:simulationtools}. 
Within section~\ref{sec:conclusion} we will outline the importance of the physics model implemented in this work in the light of our attendant publications~\cite{RadonValidation,NuclearDecay,ECR}.


\section{Electron emission accompanying radon $\alpha$-decay}
\label{sec:alphadecay}

\begin{figure}[ht!]
 \centering
    \includegraphics[width=0.8\textwidth]{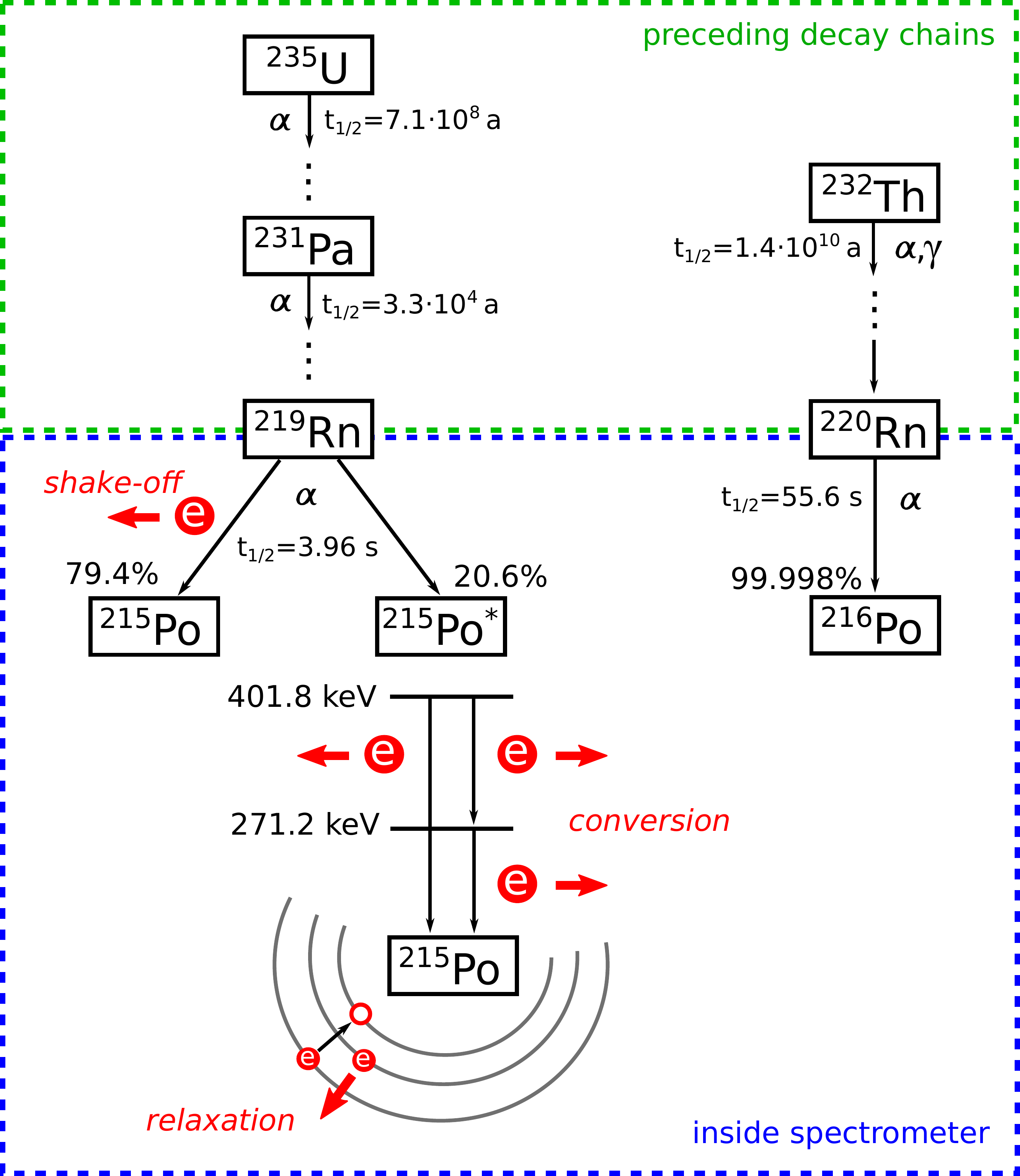}
 \caption{Top: In the KATRIN spectrometers, non-equilibrium decay chains lead to emanation of the two short-lived radon isotopes $^{219}\mathrm{Rn}$ and $^{220}\mathrm{Rn}$. Bottom: radon $\alpha$-decay processes inside the spectrometer and subsequent electron emission processes resulting from shake-off (both isotopes), conversion (mainly $^{219}\mathrm{Rn}$) and shell relaxation (following conversion and shake-off processes).}
 \label{fig:DecayChain}
\end{figure}

An essential design feature of high-resolution tritium $\beta$-spectroscopy by a MAC-E filter is an excellent UHV in the pressure range $p\leq10^{-10}$~mbar, so that background-generating ionization processes of $\beta$-decay electrons during the filter process are minimized. 
In the case of the KATRIN spectrometers, this is achieved by non-evaporable getter (NEG) strips totalling a length of 3~km in the main spectrometer and 100~m in the \prespectrometer{}. 
As shown in~\cite{Fraenkle}, the large surface of the porous NEG strips gives rise to emanation of radon atoms associated with the primordial $^{232}\mathrm{Th}$, $^{235}\mathrm{U}$ and $^{238}\mathrm{U}$ decay chains (see figure~\ref{fig:DecayChain}). 
Furthermore, the large stainless steel surface of the spectrometer vessel (main spectrometer: $650~\text{m}^{2}$, \prespectrometer{}: $25~\text{m}^{2}$) and auxiliary equipment attached to it also contributes to radon emanation due to small quantities of radon progenitors contained near the surface. 

Due to its long half-life ($t_{1/2}(^{222}\mathrm{Rn})~=~3.82~\mathrm{d}$~\cite{Nudat}), compared to the pump out time of radon in the KATRIN spectrometers ($t_{\text{prespec}}\approx25$~s, $t_{\text{mainspec}}\approx360$~s), the isotope $^{222}\mathrm{Rn}$ is essentially being pumped out of the spectrometer before it decays.
Therefore, its background contribution can be neglected. 
The short-lived isotopes $^{219}\mathrm{Rn}$ ($t_{1/2}~=~3.96~\mathrm{s}$) and $^{220}\mathrm{Rn}$ ($t_{1/2}~=~55.6~\mathrm{s}$), however, will $\alpha$-decay uniformly over the entire spectrometer volume ($V_{\text{prespec}}=8.5~\text{m}^{3}$, $V_{\text{mainspec}}=1250~\text{m}^{3}$) to their respective daughter nuclei $^{215}\mathrm{Po}$ and $^{216}\mathrm{Po}$ (see figure~\ref{fig:DecayChain}).
The important fact for our investigations here is that all $\alpha$-decays are accompanied by the emission of atomic shell electrons from the eV up to the multi-keV scale (the $\alpha$-particle as well as X-ray fluorescence photons are of no interest for our background studies, see~\cite{NuclearDecay}).
If these electrons are emitted into the sensitive volume of the spectrometer, they can contribute significantly to the background rate via secondary processes. 

There are various processes which can result in the emission of up to more than a dozen electrons in a single $\alpha$-decay. 
If the $\alpha$-decay populates an excited level of the daughter nucleus, the process of internal conversion, as described in section~\ref{sec:conversion}, will result in the emission of electrons with energies of up to several hundreds of~keV. 
Also, the $\alpha$-particle itself can interact with electrons of the inner atomic shells, leading to so called shake-off processes, detailed in section~\ref{sec:shakeoff}. 
Both processes produce vacancies in the electron shells. 
These are successively filled by atomic relaxation processes, which are the focus of section~\ref{sec:relaxation}. 
Finally, the shell reorganization process of outer shell electrons is described in~\ref{sec:shellreorg}.

\subsection{Internal conversion}
\label{sec:conversion}

In an internal conversion (IC) process the excited level of the daughter nucleus, which is populated by the $\alpha$-decay process, interacts electromagnetically with an inner-shell electron, which thus is emitted from the atom. 
As the IC process is competing with radiative processes, it is only dominant for heavy nuclei ($P(IC)\propto Z^{3}$)~\cite{Siegbahn,Theisen}, as is the case for polonium ($Z=84$). 
In addition, the probability of IC decreases for larger transition energies, so it is relevant only for low-lying levels. 
In our specific case, IC processes are thus of importance only for ${}^{219}\mathrm{Rn}\,\rightarrow\,{}^{215}\mathrm{Po}^{*}$ decays, where significant branching ratios lead to the two excited levels ($7/2^{+}$, 271.2~keV and $5/2^{+}$, 401.8~keV) shown in figure~\ref{fig:DecayChain}. 
In case of ${}^{220}\mathrm{Rn}\,\rightarrow\,{}^{216}\mathrm{Po}$ decays, the even-even nucleon configuration of the ${}^{216}$Po daughter creates a paucity of low-lying excited states, implying that IC processes following $\alpha$-decays of ${}^{220}$Rn are exceedingly rare processes.

In an IC process, an inner-shell electron with binding energy $E_{\mathrm{b}}$ is emitted into the continuum with a kinetic energy of
\begin{equation}
 E_{\mathrm{kin}}=E^{*}-E_{\mathrm{b}},
\label{equ:ICEnergy}
\end{equation}
where $E^{*}$ denotes the excitation energy of the nucleus. For our specific case of ${}^{215,216}\mathrm{Po}^{*}$ daughters, conversion electron energies between about $40~\mathrm{keV}$ and $500~\mathrm{keV}$ are observed~\cite{ConversionDataRn219,ConversionDataRn220}. 

The total conversion probability amounts to about 3.3\% when integrating over all electron shells in the case of $^{215}\mathrm{Po}^{*}$. 
The probability is largest for K-shell electrons (2\%) as they are closest to the nucleus. 
Table~\ref{tab:conversion} lists the dominant electron emission probabilities and electron energies for the decay ${}^{219}\mathrm{Rn}\,\rightarrow\,{}^{215}\mathrm{Po}^{*}$~\cite{ConversionDataRn219}. 
Our model allows for consecutive IC processes in case the initial de-excitation process does not result in a ground state configuration of the polonium daughter. 
We also include the rare IC process of the decay ${}^{220}\mathrm{Rn}\,\rightarrow\,{}^{216}\mathrm{Po}^{*}$~\cite{ConversionDataRn220}, but its contribution is negligible for the investigations in~\cite{RadonValidation,NuclearDecay}. 
As mentioned above, the emission of a conversion electron leaves a vacancy in the electron shell, leading to subsequent complex atomic shell relaxation processes, which are described in sections~\ref{sec:relaxation} and~\ref{sec:shellreorg}.

\begin{table}[!ht]
\caption{The table gives an overview of the relative probabilities $P_{i}$ (per $\alpha$-decay) of the dominant IC lines and of the corresponding electron energies $E_{\mathrm{kin}}$ for ${}^{219}$Rn, as measured by~\cite{ConversionDataRn219}. The electron energy is given by eq.~(\ref{equ:ICEnergy}) and can thus be attributed to specific excited levels of energy $E^{*}$ via the known values of the binding energy $E_{\mathrm{b}}$ (K: 93.1~keV, L: 16.9~keV, M: 4.1 keV, N: 1~keV). Only electron lines with an emission probability larger than 0.05\% are given. In our model we incorporate the possibility of consecutive IC processes within a single $\alpha$-decay in case that the 401.8~keV level of ${}^{215}\mathrm{Po}^{*}$ is populated and de-excites to the 271.2~keV level.}
\centering
  \begin{tabular}{llll}
    \hline
    \hline
    $E_{\mathrm{kin}}$ [keV] & $P_{i}$ [\%] & shell & $E^{*}$ [keV]\\
    \hline
    37.5 & 0.4 & K & 130.6 \\
    113.7 & 0.13 & L & 130.6\\
    178.13 & 1.27 & K & 271.2\\
    254.29 & 0.74 & L & 271.2\\
    267.08 & 0.19 & M & 271.2\\
    270.24 & 0.064 & NP & 271.2\\
    308.71 & 0.233 & K & 401.8\\
    384.87 & 0.102 & L & 401.8\\
    \hline
    \hline
  \end{tabular} 
  \label{tab:conversion}
\end{table}

\subsection{Inner shell shake-off processes}
\label{sec:shakeoff}

A nuclear $\alpha$-decay leads to a perturbation of the atomic shells, as the electrons experience the passage of the outgoing $\alpha$-particle through the atomic orbitals, as well as the sudden change $\Delta Z= Z'-Z$ of the Coulomb potential of the nucleus (initial state: $Z=86$ for radon, final state: $Z'=84$ for polonium)~\cite{ShakeOff}. 
The impact of both processes on inner-shell (K, L, M) electrons is different than on outer-shell (N or higher) electrons due to the largely different orbital velocities. 
For inner-shell electrons, the orbital period is much larger than the orbital passage time of the $\alpha$-particle ($v_{\alpha}/v_{e}\approx0.1$, with $v_{e}$: electron orbital velocity, $v_{\alpha}$: $\alpha$-particle velocity).
For outer-shell electrons, this ratio is reversed ($v_{e}/v_{\alpha}\approx0.1$). 
Accordingly, inner shell electrons will adjust adiabatically to the sudden change of nuclear charge. 
Outer shell electrons, however, remain 'frozen' in their parent atom ground state ($6p^{6}$ for radon) and will only slowly rearrange to the daughter orbitals ($6p^{4}$ for polonium), see section~\ref{sec:shellreorg}.

For inner shells, electron shake-off (SO) is caused by the direct collision process~\cite{ShakeOff,ShellReorganization,ShellReorg}. 
In this case, the $\alpha$-particle, which has already gained 99\% of its final kinetic energy inside the mean radius of the K-shell, can exchange energy with an electron via the Coulomb interaction when passing in the vicinity of the corresponding orbital, and, consequently, it can kick out the inner shell electron into the continuum. 
The decay energy is shared between the $\alpha$-particle and the emitted electron.
The latter carries only a small fraction, usually of the same order of magnitude as the shell binding energy $E_{\mathrm{b}}$. 
Therefore, in the adiabatic transition, the shake-off process results in a continuous, steeply falling energy spectrum. 
In this work we use the parameterization of Bang and Hansteen~\cite{Bang} to determine the emission probability for a SO electron with a certain kinetic energy $E_{\mathrm{shake}}$:
\begin{equation}
 N(E_{\mathrm{shake}})=\left(\frac{E_{\mathrm{b}}}{E_{\mathrm{b}}+E_{\mathrm{shake}}}\right)^{8}.
\label{equ:shakeoff}
\end{equation}
The SO probabilities for ${}^{210}$Po have been measured~\cite{KShakeOffRadon,LMShakeOffRadon,Szucs} and calculated~\cite{ShellReorganization,Migdal} and are used as a good approximation for the ${}^{215,216}$Po isotopes which are considered here. 
The values, which are listed in table~\ref{tab:shakeoffprob}, underline the well-known fact that the ejection probability strongly increases for higher shells. 
For the M shell, only the total emission probability is listed.
In our model, we do, however, consider the 5 subshells individually, adapting the corresponding emission probabilities.
Since there was no experimental data available for the individual subshells, we used an equal distribution amongst the subshells as an approximation.

\begin{table}[!ht]
\caption{Shake-off probabilities for electrons from specific inner shells in ${}^{210}$Po, as measured by~\cite{KShakeOffRadon,LMShakeOffRadon}. }
\begin{center}
  \begin{tabular}{lr}
  \hline
  \hline
  shell & probability \\
  \hline
  K~\cite{KShakeOffRadon} & $1.6\cdot 10^{-6}$ \\
  LI~\cite{LMShakeOffRadon} & $5.1\cdot 10^{-4}$ \\
  LII~\cite{LMShakeOffRadon} & $0.6\cdot 10^{-4}$ \\
  LIII~\cite{LMShakeOffRadon} & $1.5\cdot 10^{-4}$ \\
  M~\cite{LMShakeOffRadon} & $1.8\cdot 10^{-2}$ \\
  \hline
  \hline
  \end{tabular} 
\end{center}
\label{tab:shakeoffprob}
\end{table}

\subsection{Relaxation following conversion and shake-off processes}
\label{sec:relaxation}

\begin{figure}[ht!]
 \centering
    \includegraphics[width=0.75\textwidth]{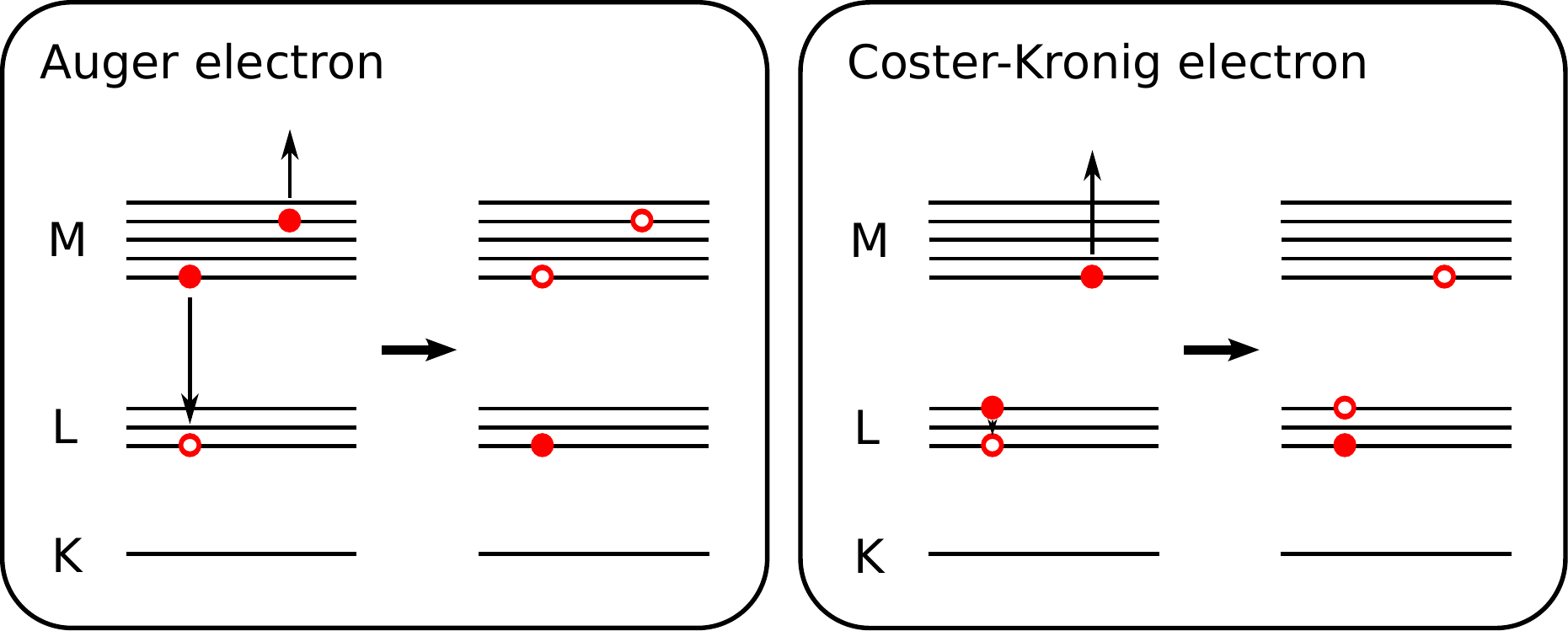}
 \caption{Sketch of a relaxation process. An inner shell vacancy is filled by an electron from an outer shell or a neighboring sub-shell, thereby releasing the corresponding binding energy difference. This energy can be transferred to another electron, which is then ejected from the atom. Depending on the origin of the electron filling the vacancy, the emitted electrons are called Auger electrons or Coster-Kronig electrons.}
 \label{fig:Auger}
\end{figure}

An electron, which is emitted via an IC or SO process, will leave a vacancy in an inner shell, as shown schematically in figure~\ref{fig:Auger}. 
As a consequence, the electron structure of the atom has to rearrange, thereby releasing binding energy. 
This can be either in the form of a fluorescence photon (radiative transition), which is of no concern for this work, or in the form of an Auger electron, if the electron filling the vacancy originates from a different shell, or a Coster-Kronig electron, in case it is emitted from a sub-shell of the same level (non-radiative transition)~\cite{Auger}. 
In case of a radiative transition, the initial vacancy is transferred to a higher atomic shell, while for non-radiative transitions the atomic shells are left with two vacancies. 
The relaxation processes then propagate up to the outermost shell.
In heavy atoms such as polonium, large cascades are observed when inner-shell vacancies are successively filled by non-radiative transitions (``Auger explosions'').
Consequently, highly charged polonium ions are created, which cannot be neutralized when propagating in the spectrometer UHV environment\footnote{The highly charged Po-ion will be neutralized when hitting the spectrometer vessel.}.

An electron emitted in a non-radiative transition will receive a distinct kinetic energy. 
In the example of figure~\ref{fig:Auger}, the Auger electron energy can, in a first approximation, be determined by
\begin{equation}
 E_{\mathrm{kin}}=(E_{\mathrm{b,L1}}-E_{\mathrm{b,M1}})-E_{\mathrm{b,M4}},
\end{equation}
where $E_{\mathrm{b,}i}$ are the binding energies of the involved shells $i$. 
In case of a radiative transition, the photon would have received the energy difference $E_{\gamma}=E_{\mathrm{b,L1}}-E_{\mathrm{b,M1}}$.
The above approximation neglects two effects~\cite{Larkins}:
\begin{itemize}
 \item A pair of holes in the atomic orbitals retains interaction energy.
 \item The relaxation of the atomic orbitals results in a lowering of the final state energy, which alters the ionization energies of electron shells containing holes.
\end{itemize}
The Auger electron energies, which are applied in our model, are indeed corrected for the aforementioned effects, using the intermediate coupling model of~\cite{Larkins}.
In the case of polonium, relaxation electron energies can reach up to about 93~keV, which approximately corresponds to the K-shell binding energy. 

Sudden changes of the atomic potentials occur during vacancy cascade development, which can lead to the emission of electrons~\cite{Pomplun}. 
However, due to their relatively low emission probabilities, this effect is only of minor importance and is neglected in our model.\\
Furthermore, as the number of vacancies in the atomic shells increases, the electron binding energies decrease which can lead to the closure of some Coster-Kronig channels, reducing the average charge state of the daughter atom~\cite{Pomplun}.
In fact, here we are not interested in obtaining an exact multiplicity distribution of emitted electrons (predominantly of exceedingly low energies $<100$~eV), because the subsequent ionization processes of high-energy IC and inner-shell SO electrons in collisions with residual gas will produce several hundred or even thousand secondary electrons due to their magnetic storage in the spectrometer.

\subsection{Atomic shell reorganization}
\label{sec:shellreorg}

In the above described processes the atomic shell of the polonium daughter is left in an excited state, and the de-excitation follows a rather complex scheme involving many different pathways within relaxation cascades. 
If the $\alpha$-decay process leaves the atomic shell unperturbed, or if the SO process of the $\alpha$-particle involves outer shell (N or higher) electrons, the relaxation processes are much less complex.
The underlying effect is that the outer-shell electron wave function cannot adjust adiabatically to the final state due to the fact that the outer-shell electron velocity is much smaller than the $\alpha$-particle velocity.
In any case the atomic system will relax to the smaller ($Z-2$) nuclear charge of the daughter nucleus.

There is a gradual transition of $\alpha$-decay processes resulting in a highly excited final state to a configuration where the atomic shell is virtually unperturbed, so that the initial state with the shell configuration of radon ($6p^{6}$) has to adjust to the ground state shell configuration of polonium ($6p^{4}$). 
In our model we treat both processes in an identical manner. 

The change in nuclear charge ($Z-2$) in the $\alpha$-decay ${}_{86}\text{Rn}\rightarrow{}_{84}\text{Po}$ results in a change $\Delta E=37.7$~keV in the total binding energy of the atomic electrons if the relativistic Hartree-Fock-Slater calculations of~\cite{HartreeFock} are used. 
$\Delta E$ is composed of the sudden energy exchange component $\Delta E_{\text{sud}}$ and the much slower rearrangement component $\Delta E_{\text{R}}$~\cite{ShellReorg}. 
As the fast inner electrons can adjust adiabatically to the effective nuclear charge reduction by rearranging to daughter orbitals, almost all of $\Delta E$ occurs suddenly ($\Delta E_{\text{sud}}$) and has to be supplied by the outgoing $\alpha$-particle, which results in an equivalent retardation. 
The remaining small fraction $\Delta E_{\text{R}}$ is retained by the atom as temporary excitation energy for the much slower shell rearrangement in the outer shells. 
We employ a scenario where the average atomic rearrangement energy $\overline{\Delta E_{\text{R}}}$ ($6p^{6}\rightarrow6p^{4}$) of about 250~eV~\cite{ShellReorg} is shared statistically by two electrons from the outermost shells. 
If their kinetic energy is larger than the polonium ionization energy for P-shell electrons (1-9~eV), they are emitted into the continuum. 
This results in a flat energy spectrum of low-energy ``shell reorganization'' electrons.

As the probability for inner shell SO (see table~\ref{tab:shakeoffprob}) and IC (see table~\ref{tab:conversion}) is not dominating, the above described atomic shell reorganization (SR) in the ground state configuration is the most frequent electron emission process accompanying $\alpha$-decay. 
If, however, the electron shell in the final state is an excited state, caused either by IC or by inner shell SO, we calculate the full atomic relaxation, which will be described in detail in section~\ref{sec:generation}.


\section{Implementation into the simulation software}
\label{sec:simulationtools}

To study the event topologies of electrons from the $\alpha$-decay of ${}^{219,220}$Rn atoms, and to estimate background rates and characteristics due to their subsequent trapping (for details, see~\cite{RadonValidation,NuclearDecay}), a detailed code for particle trajectory calculations in the complex electromagnetic field configuration of the KATRIN spectrometer is required. 
This challenging task is met by the KATRIN simulation package \textsc{Kassiopeia}~\cite{Kassiopeia}, which allows to track electrons over long periods of time with machine precision. 
For the purpose of this work a Monte Carlo event generator to describe electron emission following ${}^{219,220}$Rn $\alpha$-decay was developed and is described in section~\ref{sec:generation}. 
Section~\ref{sec:output} then gives a short overview of the output of this generator.

\subsection{The radon event generator}
\label{sec:generation}

The detailed physical model for signal events and background processes is implemented into the \textsc{Kassiopeia} package via MC-based event generators.
For the investigations of this paper, a radon background generator was developed to describe the processes accompanying the initial radon $\alpha$-decay, such as internal conversion (IC), shake-off (SO), relaxation (RX) and shell reorganization (SR) which are described in the previous section~\ref{sec:alphadecay}.

\begin{figure}[ht!]
 \centering
    \includegraphics[width=0.8\textwidth]{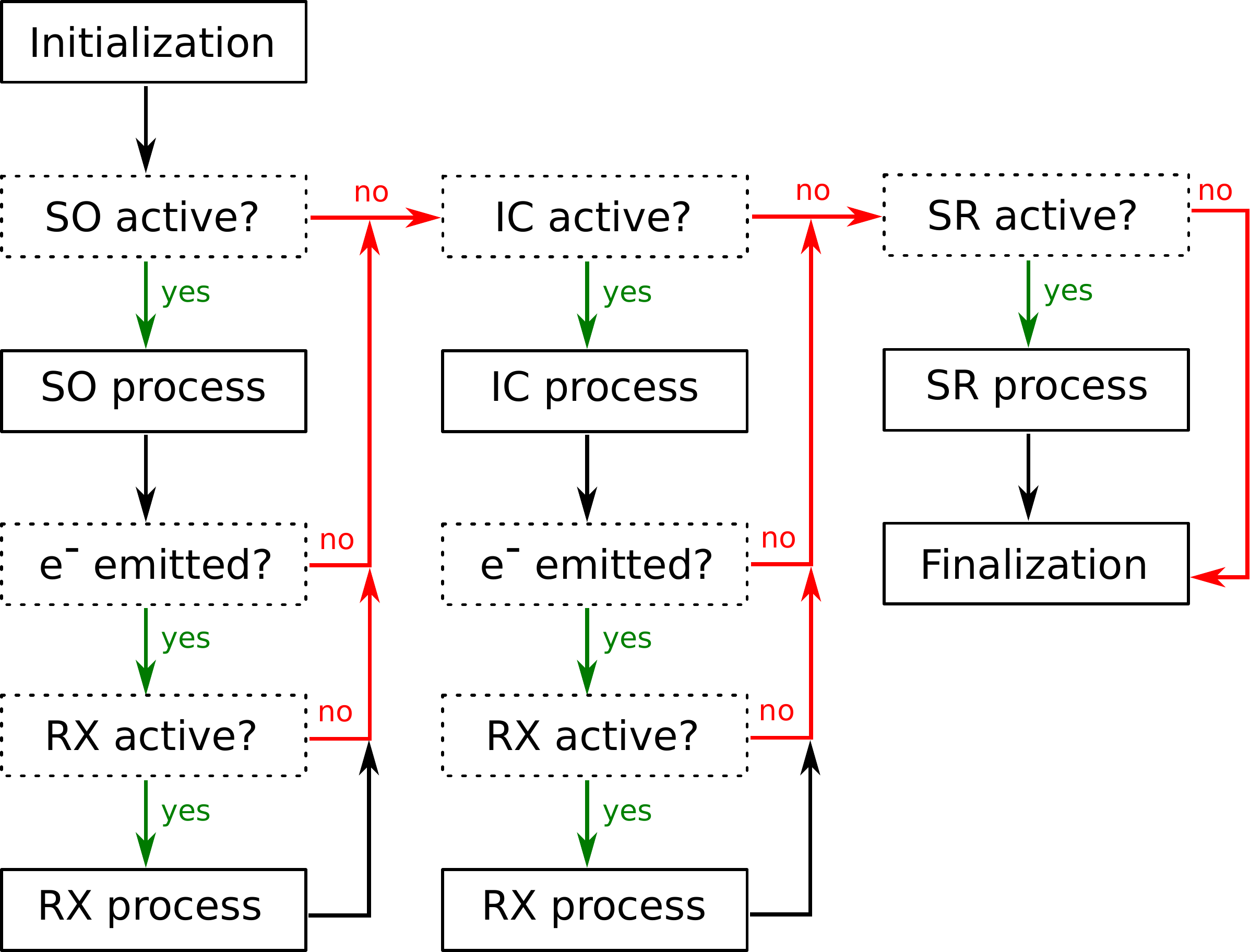}
 \caption{Event generator flowchart: After initialization of the generator, the different physical processes (SO: shake-off, IC: internal conversion, RX: relaxation, SR: shell reorganization), represented as solid boxes, are processed according to the model, which is described in more detail in the main text. The user has the possibility to configure the generator, e.g. turn on or off certain processes to study specific aspects. Corresponding decision points are given in dashed boxes.}
 \label{fig:GeneratorFlow}
\end{figure}

Figure~\ref{fig:GeneratorFlow} shows a flowchart of the radon event generator.

The simulation can be configured by the user to study the impact of different processes on the background. The following choices are available (options in brackets):
\begin{itemize}
 \item activate/deactivate individual physical processes (SO, IC, SR, RX)
 \item enforce physical processes (SO, IC)
 \item select radon isotope (219, 220)
\end{itemize}
During initialization, all data files required for the computation of the various physical processes for a specific isotope are read in.
Enforcing SO and IC processes can be useful because they are rather rare (up to few \%).
If this option is enabled, it is assured that the according processes are executed within every generated event by scaling up the emission probabilities of the individual shells until their sum totals 100\%.

The first physical process to be carried out (if activated by the user) is the SO process, as it is directly caused by the passage of the $\alpha$-particle through the atomic shell. 
At first a random number is generated by the ROOT TRandom3 routine, which is based on the Mersenne Twistor algorithm~\cite{RandomNr}. 
The SO subroutine then uses the generated random number to initiate an SO process with the corresponding probabilities for the individual (sub-)shells. 
Consequently, this can in some rare cases result in the emission of multiple SO electrons~\cite{ShakeOff}.
For the determination of the SO electron energy the acceptance-rejection method~\cite{AccRej} is applied to eq.~(\ref{equ:shakeoff}).

In case of one or more SO electrons being emitted, the RX process will be executed (if activated by the user). 
In our routine we employ the Monte Carlo technique~\cite{Pomplun2} to simulate the highly complex pathways of an initial single vacancy, where a large number of intermediate electron shell configurations is being involved. 
In a first step, we use the fluorescence yield $\omega_{i}$ and the Auger yield $\alpha_{i}$ of the shell $i$ which is under investigation to determine the transition type. 
For K- and L-shell vacancies the data of~\cite{Chen}, and for M- and N-shell vacancies those of~\cite{McGuireM,McGuireN} is used. 
If a radiative transition is diced, the vacancy is simply transferred to a higher shell, where the new vacancy is determined from the available final states according to their relative probabilities. 
Non-radiative transitions up to and including the M-shell result in two vacancies, while several vacancies can be created by N-shell vacancy de-excitation due to super-Coster-Kronig transitions, i.e.~all transitions happen within the N-shell. 
The described process is repeated until all vacancies reach the outer O- or P-shells or until no further de-excitations are energetically possible. 
As we do not take into account small modifications of the energies of electron shells due to the actual relaxation process, the de-excitations result in discrete energy lines.

After the RX process was completed, or if, initially, the SO process was deactivated, the IC process is performed (if activated by the user). 
This specific ordering is justified by the fact that shell relaxation processes are completed on a much faster time scale ($10^{-15}$~s)~\cite{Drescher} than internal conversion processes ($10^{-12}$~s)~\cite{ICTime}. 
As in the SO subroutine, a random number is used to initiate an IC process with the correct probability.
Because the excited nucleus can decay into an intermediate energy state instead of the ground state, this has to be taken into account by allowing consecutive IC processes (so called double conversion~\cite{ShellReorg}).
The interrelated energy levels are marked as such in the input file, which allows for reliable bookkeeping of the involved states.
The IC electron energy depends solely on the decaying nuclear state and the binding energy of the emitted electron, resulting in discrete IC lines.

The final process to be carried out is the SR process.
At first, the SR subroutine checks if any SO or IC processes occurred previously.
If this is the case, the SR process is skipped because the Po daughter has already relaxed via the above mentioned processes. 
Otherwise, an unperturbed shell is assumed, which results in the excitation of two electrons, statistically sharing the shell reorganization energy of $\Delta \overline{E_{\text{R}}}=250$~eV (shell binding energy to be deducted).
These electrons are actually only emitted from the atom if their energy exceeds the outer shell binding energy of about 1~eV (first ionization) or 9~eV (second ionization). 

In the final step of the event generation, all electrons generated by a single $\alpha$-decay are passed to the particle tracking part of the \textsc{Kassiopeia} simulation software.

\subsection{Generator output}
\label{sec:output}

\begin{figure}[ht!]
 \centering
    \includegraphics[width=\textwidth]{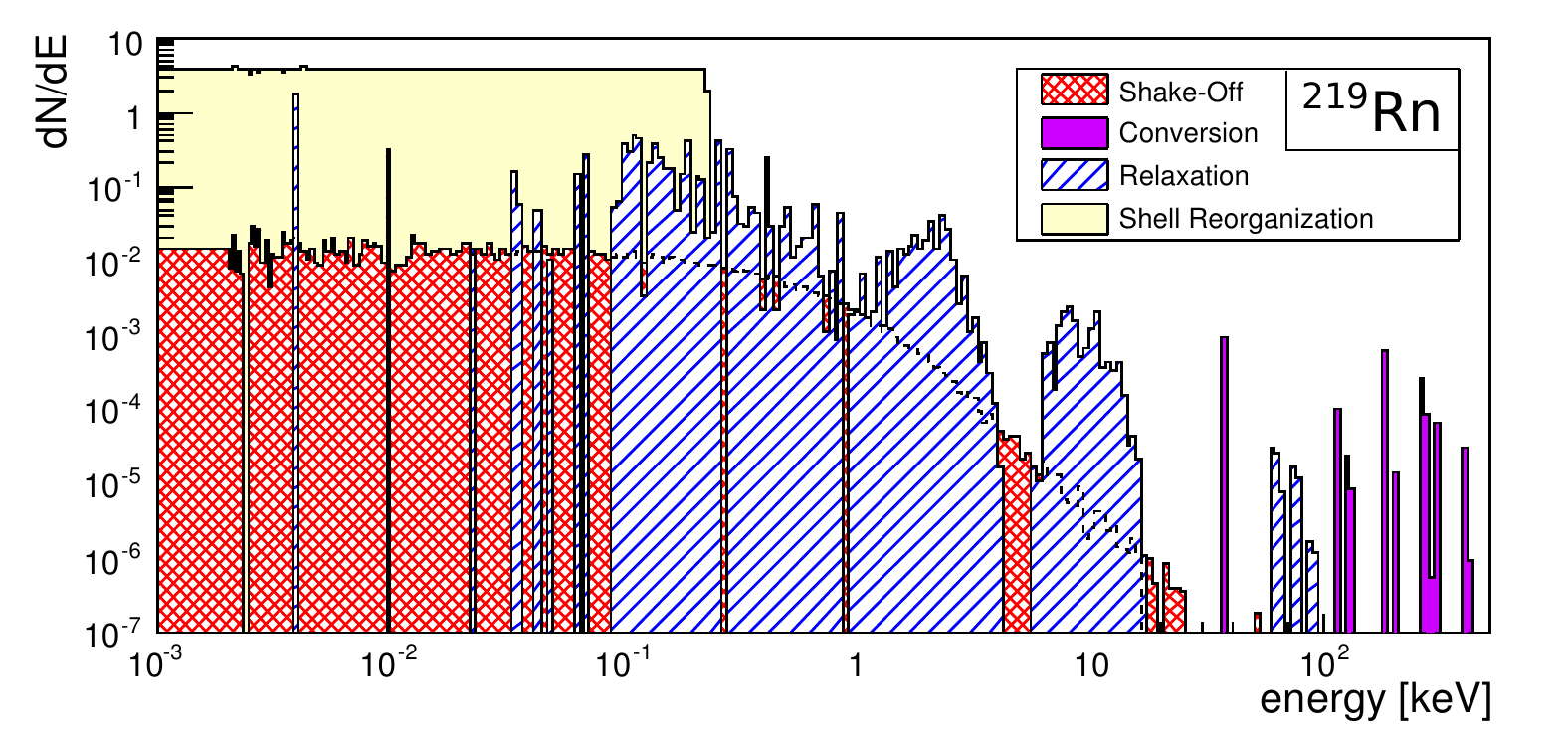}\\
    \includegraphics[width=\textwidth]{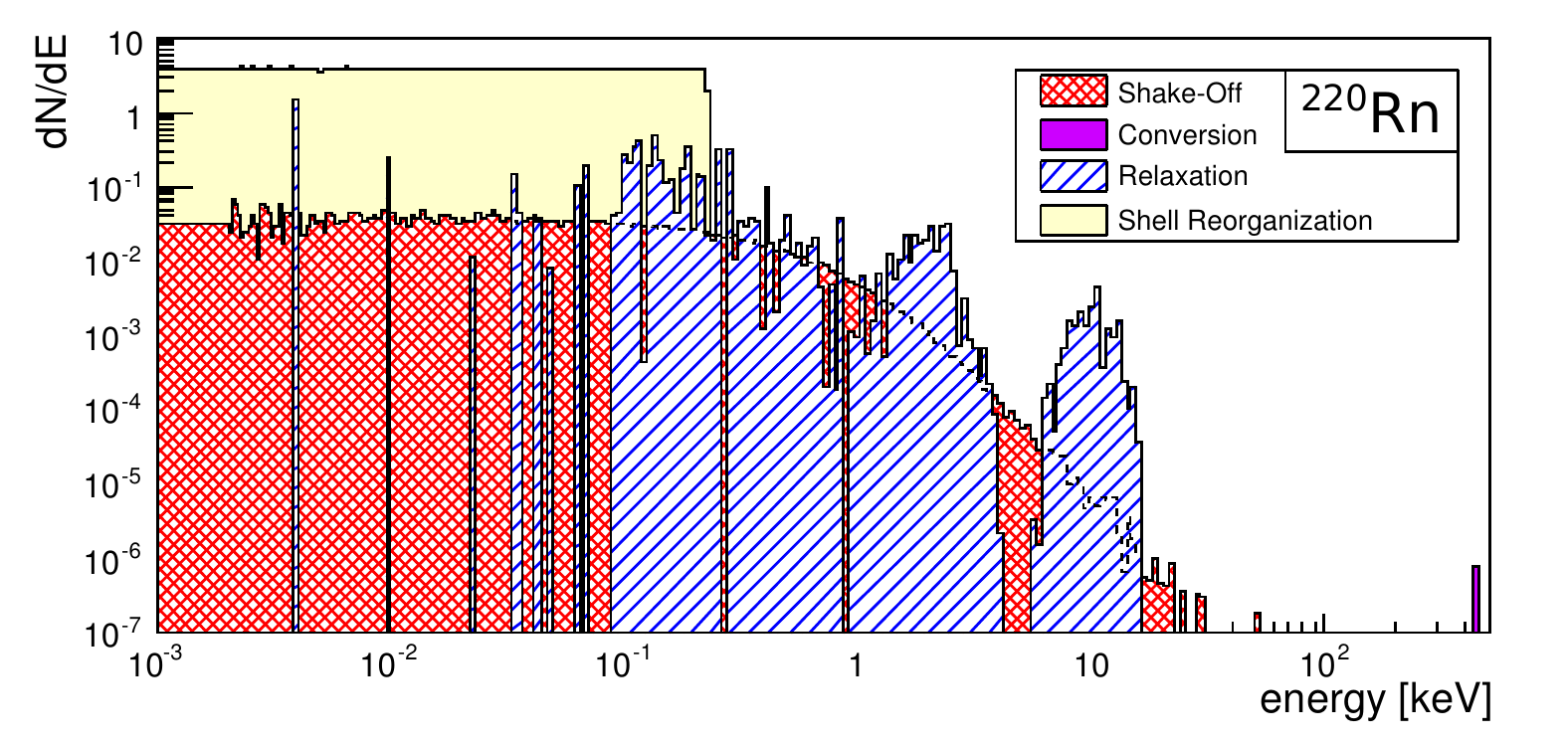}
 \caption{Event generator: energy spectra of IC, inner-shell SO, relaxation and SR electrons for the case of ${}^{219}\mathrm{Rn}\,\rightarrow\,{}^{215}\mathrm{Po}$ (top) and ${}^{220}\mathrm{Rn}\,\rightarrow\,{}^{216}\mathrm{Po}$ (bottom) $\alpha$-decay. SR electrons, which originate from unperturbed atomic shell relaxation, are distinguished from K-, L- and M-shell SO electrons. The electron energy axis is subdivided into 250 intervals between 1~eV and 500~keV with logarithmically increasing bin size.}
 \label{fig:Rn219220Gen}
\end{figure}

Figure~\ref{fig:Rn219220Gen} shows the energy spectra and energy-dependent emission probabilities as obtained with our event generator.
The discrete IC and relaxation lines, as well as the higher-order potential dependence of the SO spectrum can be clearly identified. 
The flat energy spectrum dominating the low-energy part is originating from our model of SR electrons in case of negligible atomic shell perturbation ($6p^{6}\rightarrow6p^{4}$), as the two electrons statistically share an energy of 230-250~eV.
Due to their identical nuclear charge, the inner-shell SO and SR contributions of the two polonium isotopes are assumed to be identical. 
The SO probability is negligible for the inner K-shell. 
Therefore, the low-energy part of the relaxation spectrum results mainly from L-shell (or higher) SO, and hence reaches up to about 17~keV (L-shell binding energy). 
As stated above, the IC process is of importance only in the decay ${}^{219}\mathrm{Rn}\,\rightarrow\,{}^{215}\mathrm{Po}^{*}$. 
In this case, there is a high probability for vacancies in the inner K-shell, leading to the high-energy part (up to about 90~keV) of the relaxation spectrum.

The total probability for electron emission by a specific process can be obtained by integration over the whole energy spectrum. 
The corresponding results are summarized in table~\ref{tab:GeneratorProb}.

\begin{table}[!ht]
\caption{Electron emission probabilities $P$ per decay, depending on the emission process, based on the ${}^{219}$Rn and ${}^{220}$Rn event generators of this work. $P>1$ implies that more than one electron can be emitted per decay.}
\begin{center}
  \begin{tabular}{lllll}
  \hline
  \hline
  process &  & P (${}^{219}$Rn) &  &P (${}^{220}$Rn)\\
  \hline
  inner-shell SO &  & $2.08\cdot 10^{-2}$ &  & $2.15\cdot 10^{-2}$\\
  IC &  & $3.31\cdot 10^{-2}$ &  & $5.0\cdot 10^{-5}$\\
  relaxation &  & $2.29\cdot 10^{-1}$ &  & $6.81\cdot 10^{-2}$\\
  SR &  & $1.89$ &  & $1.96$\\
  \hline
  \hline
  \end{tabular} 
\end{center}
\label{tab:GeneratorProb}
\end{table}

\subsection{Initial test of the model}

Due to the complex nature of the response of the atomic shells during and after the emission of an $\alpha$-particle, it is of vital importance to compare the present model with independent measurements. 
Generic parameters for comparison are:\\
$(i)$ the final charge state of the daughter atom, because it is highly sensitive to a correct description of processes such as atomic relaxation, and\\
$(ii)$ the electron energy spectrum in the multi-keV range, which can be estimated from the number of secondary electrons produced in the electrostatic spectrometer~\cite{RadonValidation}.\\
\begin{figure}[ht!]
 \centering
    \includegraphics[width=0.7\textwidth]{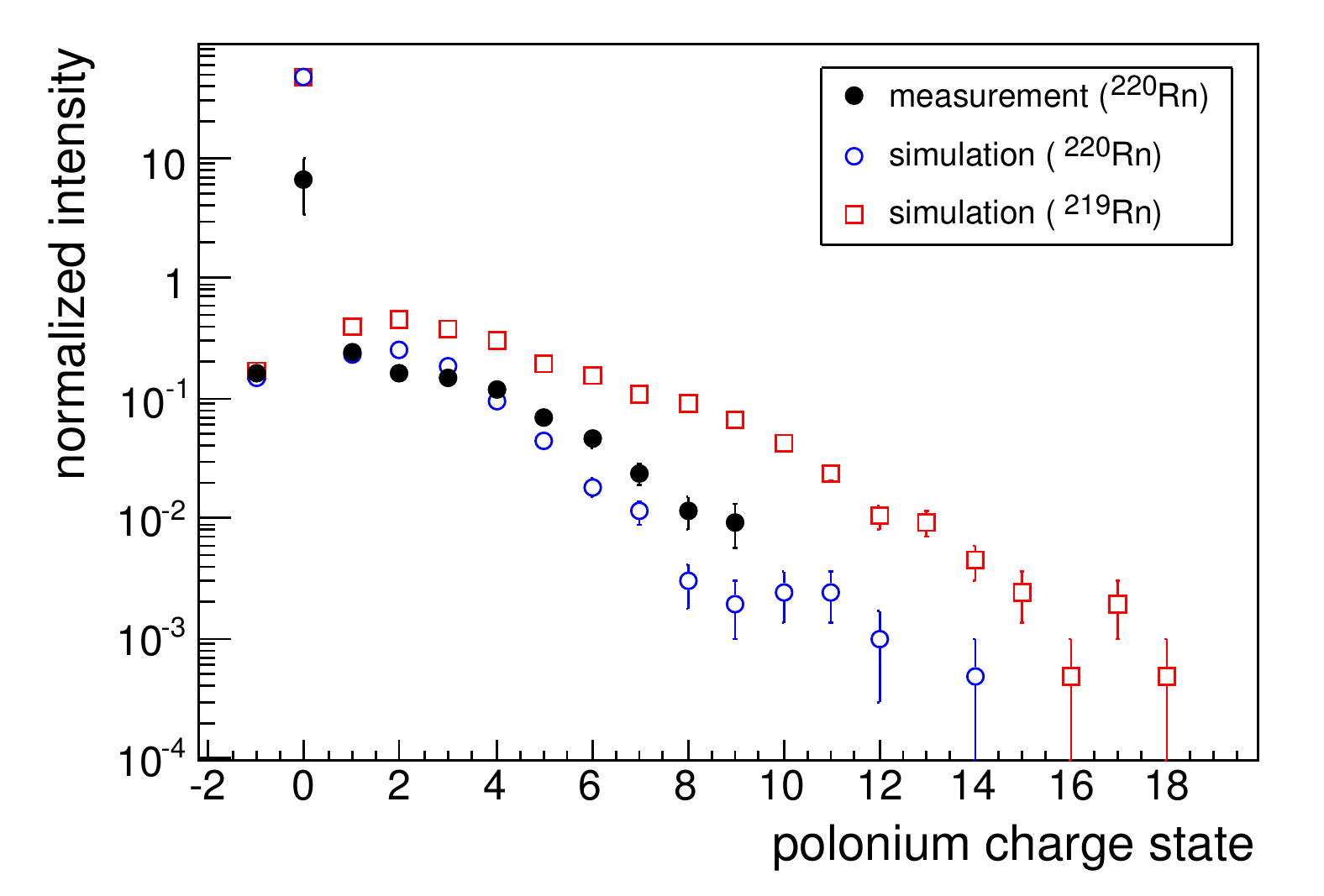}
 \caption{Charge states of ${}^{216}$Po (daughter of ${}^{220}$Rn) and ${}^{215}$Po (daughter of ${}^{219}$Rn), as obtained with our generator. The simulation is compared to an independent measurement of the ${}^{216}$Po charge state~\cite{Szucs}. The values obtained with the ${}^{220}$Rn generator were normalized to the overall rate of non-zero charge states as the experimental precision for the detection of neutral daughter atoms was rather limited in~\cite{Szucs}. For the ${}^{219}$Rn results, the same normalization constant was used to emphasize the difference between the two isotopes.}
 \label{fig:ChargeState}
\end{figure}
Figure~\ref{fig:ChargeState} shows the polonium charge state following ${}^{219}$Rn and ${}^{220}$Rn decays as obtained with the generator of this work, in comparison to the independent measurement reported in~\cite{Szucs}. 
There is good agreement between measured and simulated frequencies of occurrence of different ${}^{216}$Po charge states, which underlines the basic validity of our model.
We attribute the deviations occurring at charge states $\ge~6$ to the complex nature of atomic relaxation paths, which can only be approximated for the outer shell electrons in case of a large initial perturbation of the atomic shell system.

As can be seen in figure~\ref{fig:ChargeState}, the majority of events consists of two low energetic SR electrons. 
When comparing the experimental data of~\cite{Szucs} with our Monte Carlo generator we note that the detection of neutral daughter states, as outlined in~\cite{Szucs}, was rather challenging.
We thus ascribe the discrepancy to experimental difficulties in assessing the efficiency in detecting neutral atoms after $\alpha$-decay. 

The significant difference between the two simulated isotopes in electron multiplicities, and correspondingly in the charge distribution of the daughter ion, is due to IC processes in the case of ${}^{219}\text{Rn}\rightarrow{}^{215}\text{Po}^{*}$. 
As they are emitted from inner shells, highly charged final states result from complex relaxation cascades.

\begin{figure}[ht!]
 \centering
    \includegraphics[width=0.9\textwidth]{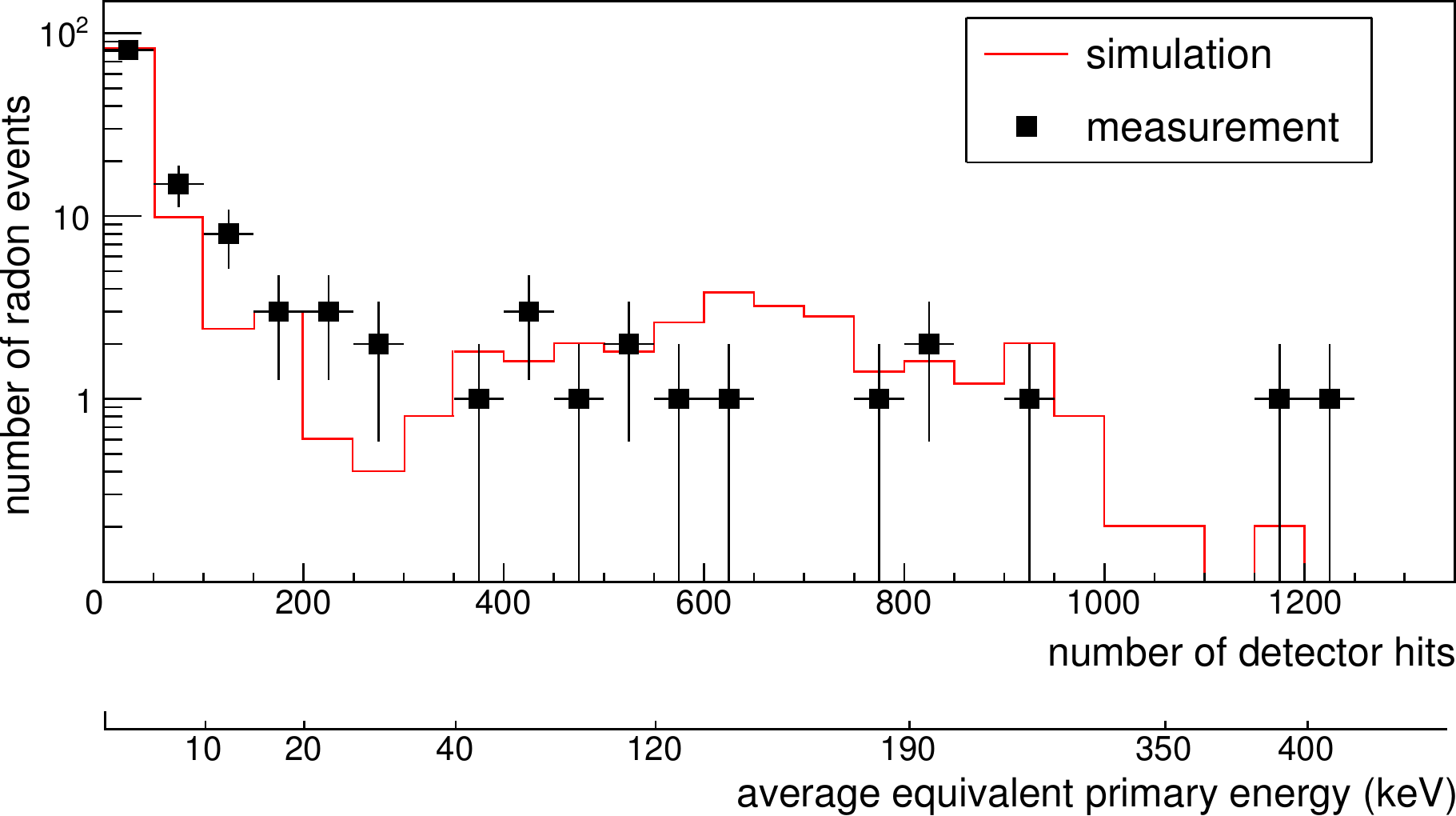}
 \caption{Comparison of measured and simulated numbers of detector hits produced by individual radon $\alpha$-decay events within the KATRIN pre-spectrometer. An equivalent energy scale can be reconstructed when using average energy losses due to scattering and synchrotron radiation~\cite{RadonValidation}. The non-linearity of the energy scale results from the decreasing scattering cross section for higher energies in combination with linearly increasing synchrotron losses.}
 \label{fig:NDet}
\end{figure}

The second important parameter which is of key importance to validate our model is the energy spectrum of the emitted electrons in the multi-keV range.
In an electrostatic spectrometer of the MAC-E filter type, this parameter cannot be measured directly, as electrons are trapped over long periods of time~\cite{Fraenkle,NuclearDecay,RadonValidation,PhDFraenkle,PhDMertens,PhDNancy}.
However, an indirect method to assess the energy of stored multi-keV electrons is to make use of their subsequent cooling via ionization of residual gas and to count the number of produced secondary electrons in a detector.
A single radon $\alpha$-decay can lead to a large number of detector hits $N_{\text{det}}$ (up to 1500 hits corresponding to a single event were observed at the KATRIN pre-spectrometer).
There is a good correlation between primary electron energy (shown in fig.~\ref{fig:Rn219220Gen}) and $N_{\text{det}}$, which is, however, not strictly linear due to competing energy losses by synchrotron radiation and due to non-adiabatic effects at higher energies.
In fig.~\ref{fig:NDet} we display the number of detector hits following single radon $\alpha$-decays in the KATRIN pre-spectrometer in an experimental configuration where ionizing collisions with residual gas (Ar at $p=2\cdot10^{-9}$~mbar) were maximized at the expense of synchrotron losses.
The measured spectrum is compared to corresponding Monte-Carlo simulations with the radon generator of this work.
There is good agreement between experimental data and MC simulation, taking into account the limited number of radon $\alpha$-decays (127 events) accumulated over a measuring period of about 500 hours.
The simulation reproduces the main features of the measured distribution: a large number of Rn-events with rather few detector counts, caused by the low-energy plateau of the shake-off events, and a steep decrease (tail of the shake-off spectrum) towards a flat plateau of very few events featuring a large number of detector hits (caused by conversion electrons).

A thorough understanding of radon-induced background is crucial for a successful neutrino mass determination with the KATRIN experiment.
Therefore, further detailed background studies comparing measurements and simulations with different experimental conditions were carried out in~\cite{RadonValidation}.
A combination of the model of this work with precise electron trajectory calculations provides the necessary thorough understanding in order to reduce the background level below the required limit.


\section{Conclusions}
\label{sec:conclusion}

In the course of this work we have developed for the first time a comprehensive and detailed model of electron emission processes following the $\alpha$-decays of the two radon isotopes ${}^{219}$Rn and ${}^{220}$Rn. 
These investigations were motivated by our earlier observations, reported in~\cite{Fraenkle}, of periods with significantly enhanced background rates at the KATRIN \prespectrometer{} measurements.

The background model incorporates various processes such as internal conversion, shake-off and relaxation of the atomic shells during or after the $\alpha$-emission. 
The resulting electron energies cover a wide range between a few~eV up to several hundred~keV, and involve highly charged polonium daughter ions. 
Our model successfully reproduces experimentally observed polonium charge multiplicities as well as electron energies in the multi-keV range, as shown in this work.
Further experimental validation of our complete physics model was performed in a separate work~\cite{RadonValidation}, where the background behavior observed within test measurements at the KATRIN \prespectrometer{} could be well described.

It is only by developing and by validating detailed models of background processes that the KATRIN experiment can realize its full physics potential in measuring the absolute mass scale of neutrinos.

\section*{Acknowledgement}
 
This work has been supported in parts by the Bundesministerium f{\"u}r Bildung und Forschung (BMBF) with project number 05A08VK2 and the Deutsche Forschungsgemeinschaft (DFG) via Transregio 27 ``Neutrinos and beyond''. We also would like to thank the Karlsruhe House of Young Scientists (KHYS) of KIT for their support (S.G., S.M., N.W.).



\bibliographystyle{unsrt}

\section*{References}

\begin{thebibliography}{10}
\expandafter\ifx\csname url\endcsname\relax
  \def\url#1{\texttt{#1}}\fi
\expandafter\ifx\csname urlprefix\endcsname\relax\def\urlprefix{URL }\fi
\expandafter\ifx\csname href\endcsname\relax
  \def\href#1#2{#2} \def\path#1{#1}\fi


\bibitem{DesignReport}
``{KATRIN Design Report (FZKA Report 7090)},'' tech. rep.,\url{ http://www.katrin.kit.edu/}, KIT, 2004.

\bibitem{Review}
G.~Drexlin {\em et~al.}, ``{Current Direct Neutrino Mass Experiments},'' 
  {\em Adv. in High Energy Phys.}, vol.~2013, ID 293986, 2013.

\bibitem{Fraenkle}
F.~Fr{\"a}nkle {\em et~al.}, ``{Radon induced background processes in the KATRIN pre-spectrometer},'' 
  {\em Astropart. Phys.}, vol.~35, no.~3, pp.~128--134, 2011.

\bibitem{MACE}
G.~Beamson, H.~Q. Porter, and D.~W. Turner, ``{The collimating and magnifying properties of a superconductiong field photoelectron spectrometer},'' 
  {\em J.  Phys. E}, vol.~13, no.~64, 1980.

\bibitem{MACE1}
V.~M. Lobashev and P.~E. Spivak, ``{A method for measuring the electron antineutrino rest mass},''
  {\em Nucl. Instrum. Meth. A}, vol.~240, no.~2, pp.~305--310, 1985.

\bibitem{MACE2}
A.~Picard {\em et~al.}, ``{A solenoid retarding spectrometer with high resolution and transmission for keV electrons},'' 
  {\em Nucl. Instrum. Meth. B}, vol.~63, no.~3, pp.~345--358, 1992.

\bibitem{Vacuum}
J.~Wolf, ``Size matters: The vacuum system of the Katrin neutrino experiment,''
  {\em Journal of the Vacuum Society of Japan}, vol.~52, pp.~278--284, 2009.

\bibitem{MagneticMirror}
H.~Higaki, K.~Ito, K.~Kira, and H.~Okamoto, ``{Electrons Confined with an Axially Symmetric Magnetic Mirror Field},'' 
  {\em AIP Conference Proceedings}, vol.~1037, no.~1, pp.~106--114, 2008.

\bibitem{NuclearDecay}
S.~Mertens {\em et~al.}, ``{Background due to stored electrons following nuclear decays at the KATRIN experiment},'' 
  {\em Astropart. Phys.}, vol.~41, no.~52, 2013.

\bibitem{RadonValidation}
N.~Wandkowsky {\em et~al.}, ``{Validation of a model for Radon-induced background processes in electrostatic spectrometers},'' 
  {\em submitted to J. Phys. G: Nucl. Partic.}, 2013.

\bibitem{PhDFraenkle}
{F.~Fr{\"a}nkle}, {\em {Background Investigations of the KATRIN Pre-Spectrometer}}.
\newblock PhD thesis, Karlsruhe Institute of Technology (KIT), 2010.

\bibitem{PhDMertens}
{S.~Mertens}, {\em {Study of Background Processes in the Electrostatic Spectrometers of the KATRIN Experiment}}.
\newblock PhD thesis, Karlsruhe Institute of Technology (KIT), 2012.

\bibitem{PhDNancy}
{N.~Wandkowsky}, {\em {PhD thesis in preparation}}.
\newblock Karlsruhe Institute of Technologie (KIT), 2013.

\bibitem{ECR}
S.~Mertens {\em et~al.}, ``{Stochastic Heating by ECR as a Novel Means of Background Reduction in the KATRIN spectrometers},'' 
  {\em JINST}, vol.~7 P08025, 2012.

\bibitem{Nudat}
A.~Sonzogni, ``{Interactive Chart of Nuclides},''
\newblock National Nuclear Data Center: Brookhaven National Laboratory, \url{http://www.nndc.bnl.gov/chart/}.

\bibitem{Siegbahn}
K.~Siegbahn, ``{Alpha-, beta- and gamma-ray spectroscopy},'' 
  vol.~2, pp.~894, North Holland Pub. Co. Amsterdam, 1968.

\bibitem{Theisen}
Ch.~Theisen, A.~Lopez-Martens and Ch.~Bonnelle, ``{Internal conversion and summing effects in heavy-nuclei spectroscopy},'' 
  {\em Nucl. Instrum. Meth. A}, vol.~589, pp.~230--242, 2008.

\bibitem{ConversionDataRn219}
E.~Browne, ``{Nuclear Data Sheets for $A = 215, 219, 223, 227, 231$},'' 
  {\em Nuclear Data Sheets}, vol.~93, no.~4, pp.~763--1061, 2001.

\bibitem{ConversionDataRn220}
S.-C. Wu, ``{Nuclear Data Sheets for $A = 216$},'' 
  {\em Nuclear Data Sheets}, vol.~108, no.~5, pp.~1057--1092, 2007.

\bibitem{ShakeOff}
M.~S. Freedman, ``{Ionization by Nuclear Transitions},''
 {\em Summer course in atomic physics, Carry-le-Rouet, France}, Aug 1975.

\bibitem{ShellReorganization}
J.~S. Hansen, ``{Internal ionization during alpha decay: A new theoretical approach},'' 
  {\em Phys. Rev. A}, vol.~9, pp.~40--43, Jan 1974.

\bibitem{ShellReorg}
M.~S. Freedman, ``{Atomic structure effects in nuclear events},'' 
  {\em Annu. Rev. Nucl. Sci.}, vol.~24, pp.~209--248, 1974.

\bibitem{Bang}
J.~Bang and J.~M. Hansteen, ``{Coulomb deflection effects on ionization and pair-production phenomena},'' 
  {\em K. Dan. Vidensk. Selsk. Mat. - Fys. Medd.}, vol.~31, no.~13, pp.~1--43, 1959.

\bibitem{KShakeOffRadon}
M.~S. Rapaport, F.~Asaro, and I.~Perlman, ``{$K$-shell electron shake-off accompanying alpha decay},'' 
  {\em Phys. Rev. C}, vol.~11, pp.~1740--1745, May 1975.

\bibitem{LMShakeOffRadon}
M.~S. Rapaport, F.~Asaro, and I.~Perlman, ``{$M$- and $L$-shell electron shake-off accompanying alpha decay},'' 
  {\em Phys. Rev. C}, vol.~11, pp.~1746--1754, May 1975.

\bibitem{Szucs}
S.~Szucs and J.~M. Delfosse, ``{Charge Spectrum of Recoiling $^{216}\mathrm{Po}$ in the $\alpha{}$-Decay of $^{220}\mathrm{Rn}$},'' 
  {\em Phys. Rev. Lett.}, vol.~15, pp.~163--165, Jul 1965.

\bibitem{Migdal}
A.~Migdal, ``{Ionization of atoms accompanying $\alpha$- and $\beta$-decay},'' 
  {\em J. Phys. (USSR)}, vol.~4, pp.~449, 1970.

\bibitem{Auger}
E.~H.~S.~Burhop, ``{The Auger effect and other radiationless transitions},'' 
  {\em University Press Cambridge}, 1952.

\bibitem{Larkins}
F.~P.~Larkins, ``{Semiempricial Auger-electron energies for elements $10\leq Z\leq 100$},'' 
  {\em At. Data Nucl. Data Tables}, vol.~20, no.~4, pp.~311--387, 1977.

\bibitem{Pomplun}
E.~Pomplun, ``{Auger Electron Spectra - The Basic Data for Understanding the Auger Effect},'' 
  {\em Acta Oncologica}, vol.~39, no.~6, pp.~673--679, 2000.

\bibitem{HartreeFock}
C.C.~Lu {\em et~al.}, ``{Relativistic Hartree-Fock-Slater eigenvalues, radial expectation values, and potentials for atoms, $2\le Z\le 126$},'' 
  {\em Atomic Data}, vol.~3, pp.~1--131, 1971.

\bibitem{Kassiopeia}
D.~Furse {\em et~al.}, ``{KASSIOPEIA - the simulation package for the KATRIN experiment},''
\newblock to be published.

\bibitem{RandomNr}
M.~Mathsumoto and T.~Nishimura, ``{Mersenne Twistor: A 623-dimensional equidistributed uniform pseudorandom number generator},'' 
  {\em ACM T. Model. Comput. S.}, vol.~8, no.~1, pp.~3--30, 1998.

\bibitem{AccRej}
C.~Robert and G.~Casella, ``{Monte Carlo Statistical Methods},'' 
  {\em Springer Texts in Statistics}, pp.~47 ff., 2004.

\bibitem{Pomplun2}
E.~Pomplun, J.~Booz and D.~E.~Charlton, ``{A Monte Carlo Simulation of Auger Cascades},'' 
  {\em Radiation Research}, vol.~111, pp.~533--552, 1987.

\bibitem{Chen}
M.~H.~Chen and B.~Crasemann and H.~Mark, ``{Relativistic radiationless transition probabilities for atomic K- and L-shells},'' 
  {\em At. Data Nucl. Data Tables}, vol.~24, no.~1, pp.~13--37, 1979.

\bibitem{McGuireM}
E.~J.~McGuire, ``{Atomic M-Shell Coster-Kronig, Auger, and Radiative Rates and Fluorescence Yields for Ca-Th},'' 
  {\em Phys. Rev. A}, vol.~5, no.~3, pp.~1043--1047, 1972.

\bibitem{McGuireN}
E.~J.~McGuire, ``{Atomic N-Shell Coster-Kronig, Auger, and radiative rates and fluorescence Yields for $38<Z<103$},'' 
  {\em Phys. Rev. A}, vol.~9, no.~5, pp.~1840--1851, 1974.

\bibitem{Drescher}
M.~Drescher {\em et~al.}, ``{Time-resolved atomic inner-shell spectroscopy},'' 
  {\em Nature}, vol.~4198, pp.~803--807, 2002.

\bibitem{ICTime}
D.W.~McCamant, P.~Kukura and R.A.~Mathies, ``{Femtosecond Time-Resolved Stimulated Raman Spectroscopy: Application to the Ultrafast Internal Conversion in $\beta$-Carotene},'' 
  {\em J. Phys. Chem. A}, vol.~107, no.~40, pp.~8208--8214, 2003.








\end{thebibliography}

\end{document}